\documentclass[preprint,11pt]{aastex}
\usepackage{natbib}
\newcommand{\civ}{C\,{\sc iv}}

\newcommand{\ho}{$H_0$}

\newcommand{\kms}{km\,s$^{-1}$}
\newcommand{\kmsm}{km\,s$^{-1}$\,Mpc$^{-1}$}

\newcommand{\oii}{[O\,{\sc ii}]}

\newcommand{\OIII}{[O\,{\sc iii}]}

\newcommand{\CII}{C\,{\sc ii}\,$\lambda$1335}

\newcommand{\lya}{Ly$\alpha$}
\newcommand{\LyA}{Ly$\alpha$\,$\lambda$1216}

\newcommand{\NV}{N\,{\sc v}\,$\lambda$1240}

\newcommand{\OI}{O\,{\sc i}\,$\lambda$1302}
\newcommand{\OII}{[O\,{\sc ii}]\,$\lambda$3727}

\newcommand{\OVI}{O\,{\sc vi}\,$\lambda$1034}

\newcommand{\SiO}{Si\,{\sc iv}/O\,{\sc iv}]\,$\lambda$1400}

\shortauthors{HALL ET AL.}
\shorttitle{CNOC2 GRAVITATIONAL LENS CANDIDATES}
\slugcomment{Accepted to The Astronomical Journal}

\begin{document}

\title{Spectroscopic Gravitational Lens Candidates
in the CNOC2 Field Galaxy Redshift Survey}

\author{
Patrick B. Hall,\altaffilmark{1,7}
H. K. C. Yee,\altaffilmark{1,7}
Huan Lin,\altaffilmark{2,7,8}
Simon L. Morris,\altaffilmark{3,7}
Michael D. Gladders,\altaffilmark{1,7}
R. G. Carlberg,\altaffilmark{1,7}
David R. Patton,\altaffilmark{1,4,7}
Marcin Sawicki,\altaffilmark{1,5,7}
Charles W. Shepherd,\altaffilmark{1}
and Gregory D. Wirth\altaffilmark{4,6,7}
\altaffiltext{1}{Department of Astronomy, University of Toronto, 60 St. George
Street, Toronto, ON M5S~3H8, Canada; E-mail: hall@astro.utoronto.ca}
\altaffiltext{2}{Steward Observatory, The University of Arizona, Tucson, AZ 85721, USA}
\altaffiltext{3}{Dominion Astrophysical Observatory,
Herzberg Institute of Astrophysics, National Research Council, 
5071 W. Saanich Rd., Victoria, BC V8X 4M6, Canada}
\altaffiltext{4}{Department of Physics and Astronomy, University of Victoria,
PO Box 3055, Victoria, BC V8W 3P6, Canada}
\altaffiltext{5}{California Institute of Technology, Mail Stop 320-47,
Pasadena, CA 91125, USA}
\altaffiltext{6}{W. M. Keck Observatory, Kamuela, HI 96743, USA}
\altaffiltext{7}{Visiting Astronomer, Canada-France-Hawaii Telescope,
operated by the National Research Council of Canada, the Centre National
de la Recherche Scientifique de France and the University of Hawaii.}
\altaffiltext{8}{Hubble Fellow}
}

\begin{abstract}

\small

\noindent{We present five candidate gravitational lenses discovered 
spectroscopically in the Canadian Network for Observational Cosmology Field
Galaxy Redshift Survey (CNOC2), along with one found in followup observations.
Each has a secure redshift based on several features, 
plus a discrepant emission line which does not match any known or plausible
feature and is visible in multiple direct spectral images.
We identify these lines as 
\LyA\ or \OII\ emission from galaxies lensed by,
or projected onto, the CNOC2 target galaxies.
Einstein radii estimated from the candidate deflector galaxy luminosities
indicate that 
for two candidates the lines are probably \oii\ from projected $z<1$ galaxies
(consistent with the detection of H$\beta$ as well as \oii\ in one of them),
but that in the remaining four cases the lines could be Ly$\alpha$ from lensed
$z>3$ galaxies.
We estimate that only $1.9\pm0.7$ \oii-emitting galaxies are expected to
project onto target galaxies in the original CNOC2 sample, consistent with
three or four of the six candidates being true gravitational lenses.
}

\end{abstract}

\small


\section{Introduction}	\label{Intro}

Gravitational lensing of one galaxy by a foreground galaxy can provide
unique insight into the mass distribution of the deflector galaxy 
and the morphology of the source galaxy
when the lensed emission extends over a range of radii \citep{koc95,koc98}.
Radially extended emission occurs in the form of full or partial Einstein rings
which are formed when the background galaxy is of order $\sim$1~arcsec in size,
comparable in size to the inner (diamond) caustic of a typical deflector galaxy.
\citet{mel92} pointed out that a large population of optical Einstein rings
should exist if the background density of faint galaxies is high enough.
The first such (partial) optical Einstein ring was discovered in a spectroscopic
survey of $z$$\sim$0.4 early-type galaxies \citep{war96,war99},
and another has since been discovered in the same survey \citep{hew99}.
The main advantage of spectroscopic galaxy-galaxy lens searches is that the
source and deflector redshifts, and thus the basic lensing geometry,
are determined upon discovery 
(though further observations are needed to obtain the exact lens morphology
  and possibly additional confirmation of the objects' lensing natures).
In contrast, 
galaxy-galaxy lens candidates selected from direct imaging
can suffer from confusion with morphologically complex galaxies
even when high resolution images are available \citep{rat99},
and while radio-selected lenses have high-resolution discovery images,
they often suffer from faint optical counterparts which make redshifts
difficult to obtain \citep{tk00}.

In this paper we present a sample of candidate gravitational lenses identified
spectroscopically during the Canadian Network for Observational Cosmology Field
Galaxy Redshift Survey (CNOC2) and a subsequent followup project.  
Each consists of a putative {\em deflector} galaxy (the CNOC2 target galaxy) 
whose flux dominates the spectrum and yields a secure redshift,
plus a discrepant emission line from a putative {\em source} galaxy
lensed by, or projected onto, the target galaxy.
Existing broadband imaging shows no firm morphological evidence for lensed
emission, but
high-resolution narrowband imaging should reveal that the real 
gravitational lenses among these candidates are full or partial Einstein rings.  
We assume \ho=100\,\kmsm\ ($h$=1), $\Omega_M$=0.2 and $\Omega_\Lambda$=0.

\section{Data}	\label{Data}

The CNOC2 dataset is discussed in \citet{yea00} and only relevant details are
summarized here.  
$UBVR_cI_c$ imaging and spectroscopy were obtained using the
Multi-Object Spectrograph (MOS) on the Canada-France-Hawaii Telescope (CFHT).
The survey covers a total area of 5434~arcmin$^2$ 
divided among four {\em patches} on the sky, each of which is a mosaic of
17 to 19 contiguous MOS fields of size $\sim$9\arcmin$\times$8\arcmin.
Spectroscopy was obtained for two slit-masks per field (denoted A and B),
with two exposures per mask
(each typically 20 minutes long for the A masks and 40 minutes for the B masks),
east-west slits of width 1\farcs3 and minimum length 10\farcs5,
and typically $\sim$100 objects per mask.
The final rebinned 1-D spectra cover 4390 to 6292~\AA\ at 4.89~\AA/pixel
and resolution $\sim$14.8\,\AA.
	They are thus capable of detecting Ly$\alpha$ at $2.620<z<4.167$
	and \oii\ at $0.181<z<0.685$.
In addition, followup red spectroscopy was obtained in 1999 August to measure
the Balmer decrements of a subset of CNOC2 galaxies with known redshifts
\citep{mor00}.  These spectra cover 5000 to 9000~\AA\ at 5~\AA/pixel and 
resolution $\sim$17\,\AA.
Redshifts were determined by cross-correlation with a set of three templates: 
one absorption-line (spectral class Scl=2; E galaxy), 
one emission+absorption (Scl=4; Sbc galaxy), 
and one emission-dominated (Scl=5; Scd galaxy).
The choice of template to use, 
and thus the spectral class and redshift assigned to the galaxy,
was made after a visual review which usually involved examining
the direct and spectroscopic images and the final 1-D spectra.
The CNOC2 catalog used in this paper (2000 April version)
contains 6130 spectroscopically identified galaxies.

\section{Selection of Spectroscopic Lens Candidates}	\label{Selection}

\begin{figure}
\plotone{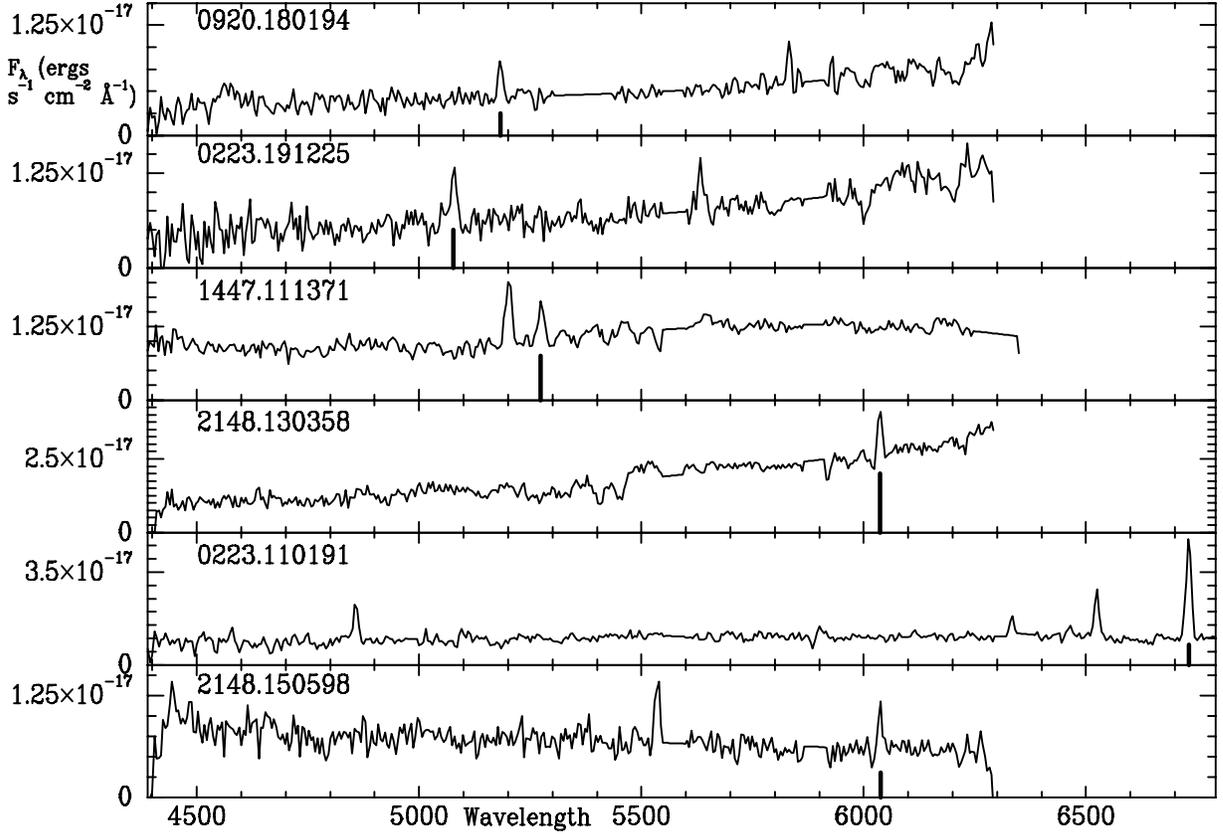} 
\caption[]{
\singlespace
\small
Observed spectra of the six candidate spectroscopic gravitational lenses.
The spectra are ordered by putative deflector galaxy redshift,
with the highest redshift at top.
The discrepant lines are marked with a vertical dash underneath the continuum.
Regions around night sky lines at 5577~\AA\ and 5892~\AA\ are interpolated over,
as well as the 5300--5430~\AA\ region in 0920.180194 which was contaminated by
zeroth-order emission from another slit.
The apparent line at 4444~\AA\ (observed) in 2148.150598 is spurious.
All available spectra for each object were coadded to improve the SNR.
The spectrum of 0223.110191 extends to redder observed wavelengths than the 
others since it includes data obtained with a different spectrograph setup
(\S\ref{Data}).
Note the different vertical flux scales.
}\label{stack}
\end{figure}

During the visual redshift assignment review for five CNOC2 targets 
and one followup target, we noted the presence of
a strong unidentified emission line discrepant from the adopted redshift.
The redshifts for these targets were determined from multiple spectral features
and have $R_{cor}\gtrsim4$,
where $R_{cor}$ is the SNR of the redshift correlation peak \citep{yec96}.
Figure~\ref{stack} shows the observed spectra of the candidate lenses
(in all our discussions we present the candidates
in decreasing order of deflector galaxy redshift).
Each candidate's ID number is its CNOC2 catalog number \citep{yea00}, consisting of a
four-digit patch code plus a six-digit field+object code after a decimal point.
All of the discrepant lines are significant detections and are present in both
exposures of each mask in which they were observed.  The possible causes
of spurious lines (including zeroth order emission, cosmic rays, spatially and
spectrally adjacent slit overlaps, and flatfield residuals) are ruled out in
each case by visual inspection of the two-dimensional (2-D) spectroscopic images.
When the spectra are examined in the rest frame of the target galaxy,
none of the discrepant lines match the wavelength
of any emission line seen in the 719 local galaxies studied by \citet{san78},
	nor the wavelength of any feature which mimics an emission line 
	as empirically observed in the CNOC2 absorption-line galaxy spectra.
Also, in the five of these six objects with more than one observed emission
line, the wavelength ratios of the lines do not match any known or reasonable
pair of UV or optical lines at any $z$ (but see \S\ref{0223.191225} below).
Thus, these objects are best described as candidate gravitational lens systems,
having one set of spectral features plus a discrepant emission line which is
almost certainly either \oii\ or Ly$\alpha$ \citep[see discussion in][]{war96}.
Table~\ref{tab_lenses} and Table~\ref{tab_phot} 
give the spectroscopic and
photometric parameters, respectively, for these candidate~lenses.

\begin{figure}
\plotone{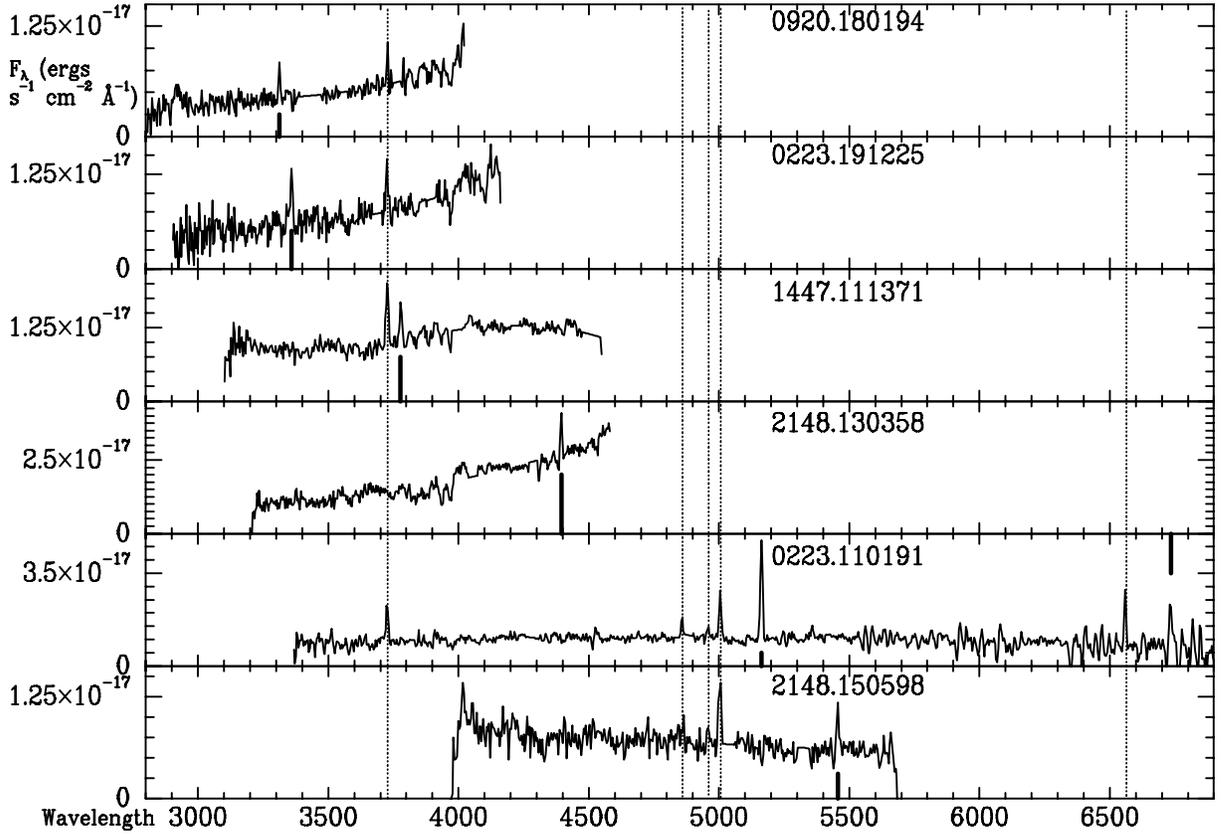} 
\caption[]{
\singlespace
\small
Spectra of the six candidate spectroscopic gravitational lenses in the
deflector galaxy rest frame.
The spectra are ordered by putative deflector galaxy redshift,
with the highest redshift at top.
The discrepant lines are marked with a vertical dash underneath the continuum.
For 0223.110191, the discrepant line is confirmed to be \oii\ by the detection
of a second discrepant line (marked with a dash above the continuum) at the
wavelength of H$\beta$ at the same redshift.  
From left to right, the dotted lines show the positions of
\OII, H$\beta$, [O\,{\sc iii}]\,4959, [O\,{\sc iii}]\,5007 and H$\alpha$.
Regions around night sky lines at 5577~\AA\ and 5892~\AA\ are interpolated over.
All available spectra for each object were coadded to improve the SNR.
The spectrum of 0223.110191 extends to redder observed wavelengths than the 
others since it includes data obtained with a different spectrograph setup
(\S\ref{Data}).
Note the different vertical flux scales.
}\label{stack2}
\end{figure}

Figure~\ref{stack2} shows the observed spectra of the candidate lenses
in the putative deflector galaxy rest frame.
Five of the six lens candidates have emission lines at
the deflector galaxy redshift in addition to the discrepant emission line.
For targets such as these where the redshift was determined from an emission
line template, it is relatively easy to spot even weak discrepant emission
lines during the visual redshift assignment review.
However, it is possible that discrepant but weak emission lines were ignored
during this step if the redshift was assigned from an absorption line template.
Thus the final 1-D spectra for all $\sim$2350 galaxies 
with redshifts assigned from the absorption line template
or the emission+absorption line template
were re-examined by eye for discrepant emission lines.
No additional candidates were found with confirmed lines
as strong as in the original five CNOC2 candidates.
Therefore we are confident that in the CNOC2 database there are no other
candidate spectroscopic lenses with discrepant lines as strong as these five 
($\geq 5~10^{-17}$ ergs cm$^{-2}$ s$^{-1}$ or equivalent width $\gtrsim10$\,\AA)
and with observed wavelengths visible in CNOC2.

It is worthwhile to note that we found only one strong candidate among objects
with purely absorption line spectra, despite the fact that such objects are
typically bulge-dominated galaxies which are more efficient lenses than 
disk-dominated galaxies.  
We feel this is probably not a significant result for several reasons.
First, our sample is only six objects.  
Second, morphologically most of them appear to be bulge-dominated
(\S\ref{Notes}).
Third, the distribution of spectral classes of our candidate deflector galaxies
reflects not only lensing efficiency but also the fraction of each spectral
class in the primary $R\leq21.5$ CNOC2 spectroscopic sample (25\% absorption
  line objects, 20\% emission+absorption and 55\% emission line).
Fourth, \citet{sch99} find \oii\ emission in one-third of a sample of $z$$<$1
galaxies classified as bulge-dominated using $HST$ images.

Assuming the discrepant line is Ly$\alpha$,
none of the candidates exhibit a significant Lyman break or significant emission
from Ly$\beta$/\OVI, \NV, \OI, \CII\ or \SiO.  
This is not surprising, since some of these emission lines are only strong in
AGN and since star-forming galaxies can have extremely strong Ly$\alpha$
relative to their continua \citep{ste00}.

\section{Discussion}	\label{Discussion}

The five candidate lenses from the original CNOC2 survey form a unique sample
from a survey with well-understood selection effects (the sixth lens was
  identified from an ongoing and thus statistically incomplete survey,
  and is excluded from the following discussion).
The selection biases for these lens candidates are:
the deflector galaxy must give a secure redshift;
the source galaxy must be at $2.620<z<4.167$ for Ly$\alpha$ 
or $0.181<z<0.685$ for \oii;
and the source galaxy must have an emission line strong enough 
to be visible.\footnote{The bias toward deflector galaxies with secure
redshifts means that a system with a source at $2.620<z<4.167$ and a deflector
at $z\gtrsim0.6$ could have been misidentified as an \oii\ emission-line
galaxy in the CNOC2 database.}  
We now attempt to account for these biases and estimate the number of projected
\oii-emitting galaxies (foreground or background to the CNOC2 target galaxies)
which could contaminate the lens sample from the original CNOC2 survey.
We defer a discussion of the expected number of lenses in the survey
to \S\ref{Possibilities}.

\subsection{Estimated Number of Projected \oii-emitting Galaxies} \label{Projections}

Since we cannot yet definitively identify the discrepant lines observed in
these objects as arising from lensed galaxies, we now consider the possibility
that they are \oii\ from galaxies seen in projection along the line of sight.
We adopt 1\farcs3$\times$1\farcs3 (4.694~10$^{-4}$ arcmin$^2$)
as the area within which a galaxy would have to be projected to produce
a discrepant line in our spectra, 
since we use 1\farcs3 wide slits and the typical seeing is better than 1\farcs0.
The spectral profiles along the slits and the imaging data on individual
objects discussed in \S\ref{Notes} rule out larger projected separations.
The expected number of chance projections in the CNOC2 sample which might
explain these objects is then just this area times the surface density
of galaxies which have properties consistent with the putative \oii\ emitters.
The imaging data discussed in \S\ref{Notes} rules out projected galaxies with
magnitudes comparable to the target galaxies.
We adopt bright magnitude limits for potential projected galaxies 2\fm5
fainter than the target galaxy (i.e. ten times lower flux).
A conservative faint magnitude limit can be estimated by assuming the
discrepant line flux comes from an object whose \oii\ rest-frame equivalent
width (REW) is 100\,\AA, the maximum value observed even in starburst galaxies
\citep[but see][]{ste00}.
The galaxy surface density in these magnitude ranges
(typically $R=22.5-24$) is 10.22~arcmin$^{-2}$ on average.
Since $0.181 \leq z \leq 0.685$ is required for \OII\ to be seen in CNOC2,
we multiply by a correction factor $f_{z}$, 
the fraction of galaxies predicted to have $0.181 \leq z \leq 0.685$.
These predictions are taken from the model of \citet{gy00c}, 
which uses extrapolated CNOC2 luminosity functions \citep{cnoc2.1}
to simultaneously fit numerous observational constraints.
On average, $f_{z} \simeq 0.32 \pm 0.12$.  
We also need to multiply by an additional correction factor $f_{\rm OII}$,
the fraction of galaxies in these magnitude and redshift ranges 
with \oii\ emission strong enough to match the observations.  
This fraction depends on magnitude since fainter galaxies 
must have higher \oii\ REW to explain the observed line flux.  
Even at our assumed bright magnitude limit, the required \oii\ REW is
$\sim$25\,\AA.  Approximately $\sim$35\% of $R\sim23$ galaxies with
$0.181 \leq z \leq 0.685$ have \oii\ REW$>$25\,\AA\ in the sample of
\citet{hog98}, but only $\sim$2\% have REW$>$100\,\AA.
We assume $f_{\rm OII}\sim0.2$ averaged over our magnitude ranges.

We use all these numbers to estimate the total number of galaxies which could
explain the observed discrepant emission lines.  We obtain
$1.9\pm0.7$ total chance projections 
using the formula $N A n f_{z} f_{\rm OII}$
with $N=6130$ CNOC2 galaxies with redshifts,
$A = 4.694~10^{-4}$ arcmin$^2$/galaxy, 
$n = 10.22$ candidate projected galaxies/arcmin$^2$, 
$f_{z} = 0.32 \pm 0.12$
and $f_{\rm OII} = 0.2$.
Using Poisson statistics with this mean, we estimate that there is only a 3\% 
chance that all five candidates from the original CNOC2 survey are projections.

A firm upper limit to the number of projected \oii-emitting galaxies can also
be made from the CNOC2 data itself.  We calculate $n f_{z} f_{\rm OII}$ 
independently by measuring, in the appropriate magnitude range $22.5<R<24$,
the surface density of all galaxies times the fraction of CNOC2 targets with
redshifts $0.181 \leq z \leq 0.685$ from emission or emission+absorption line
templates.  This method yields an estimate of $3.1\pm0.6$ total chance
projections, higher than above but within the uncertainties.  
This is expected to be an overestimate because it does not account for the
fact that some of these galaxies have weak \oii\ which would not be detected
at a significant level if the galaxy was projected atop a brighter galaxy.

\subsection{Estimated Einstein Radii}	\label{Einstein}

The cross-section for lensing scales as $\pi\theta_E^2$ \citep{bn92}.
Thus the estimated Einstein radii of our candidate deflector galaxies can be
used as a consistency check on the probability of detecting such lenses.
For a singular isothermal sphere, the Einstein radius
$\theta_E = 1\farcs455 \sigma_{225}^2 D_{LS} / D_{OS}$, where $\sigma_{225}$
is the central dark matter velocity dispersion in units of 225~\kms\ and
$D_{OS}$ and $D_{LS}$ are the angular diameter distances to the source
from the observer and the deflector (lens) galaxy, respectively \citep{koc99}.
\citet{koc94} found that for a sample of local early-type galaxies,
$\sigma_{225} = (L/L_*)^{0.24}$,
where $L_*$ is the luminosity corresponding to $M_B=-19.9+5{\rm log}(h)$.
We use these two equations to calculate $\theta_E$ for our candidate lenses
for each possible discrepant line identification for each object.
However, to ensure that our galaxies' $M_B$ values are
directly comparable to those of the \citet{koc94} galaxies, we need to make two
corrections to the $M_B$ values in Table 2 before using them in these equations.
The \citet{koc94} galaxies are local ellipticals with insignificant levels of
blue light from star formation, but our galaxies have considerable blue light,
as seen in their spectra and indicated by their spectral and SED classes.  
The lensing cross-section will
be overestimated directly in proportion to the fraction of excess blue light.
To account for this bias, we recomputed $M_B$ by normalizing to the observed
$I$ band flux and assuming an SED class of 0.38, 
equal to the earliest type observed among the candidates.
We then correct for the passive evolution observed
in the population of galaxies with early-type SEDs in CNOC2.  
This is parametrized as $M_B(0)=M_B(z)+Qz$ \citep{cnoc2.1},
with $Q=1.07$ from an analysis of the full CNOC2 sample using evolving GISSEL
models to define the galaxy SED types \citep{lea00}.
The $M_B$ now represent our best estimates of the magnitudes 
the ``old population" light in these galaxies would have at $z=0$,
and we use them to calculate $\theta_E$(Ly$\alpha$) and $\theta_E$(\oii).
The results are given in Table~\ref{tab_phot} and discussed in the next section.

\subsection{Notes on Individual Candidates: Projections or Lenses?} \label{Notes}

Since the cross-section for lensing scales as $\pi\theta_E^2$,
lenses with small $\theta_E$ are much rarer than lenses with large $\theta_E$.
Our survey is sensitive to lenses of any $\theta_E$
(unlike imaging surveys for morphologically selected lenses), but we are still
sensitive to this cross-section bias against the existence of small-$\theta_E$
lenses.  Given this bias, our estimated $\theta_E$ values are about
as large as expected for galaxies which are acting as gravitational lenses,
except for 2148.150598 and 0223.110191.  These objects have inferred masses 
too low to have a high probability of lensing.  This finding of one or two 
probable superpositions from lensing probability considerations agrees
well with the estimate of $1.9\pm0.7$ total chance projections 
(in the original CNOC2 survey) of objects which could explain
the discrepant emission lines (\S\ref{Projections}).
{\em This agreement suggests that the other candidate lenses are probably real
gravitational lenses.}  
Further confirmation is clearly necessary, but in the meantime
in this section we outline the probable natures of the individual candidates.

\begin{figure}
\plotone{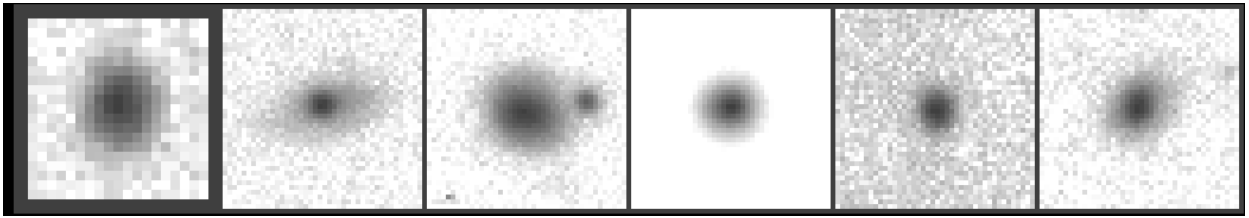} 
\caption[]{
\singlespace
\small
$R$ images of the six candidate lenses, ordered by putative deflector galaxy
redshift, with the highest redshift at left.
Each image is 8\farcs4 on a side, with East to the left and North up.
From left, the objects are 
0920.180194, 0223.191225, 1447.111371, 2148.130358, 0223.110191 and 2148.150598.
CFH12k data with 0\farcs2 pixels are shown for all objects except 0920.180194,
for which only MOS data with 0\farcs438 pixels is available.
}\label{figure}
\end{figure}

\begin{figure}
\plotone{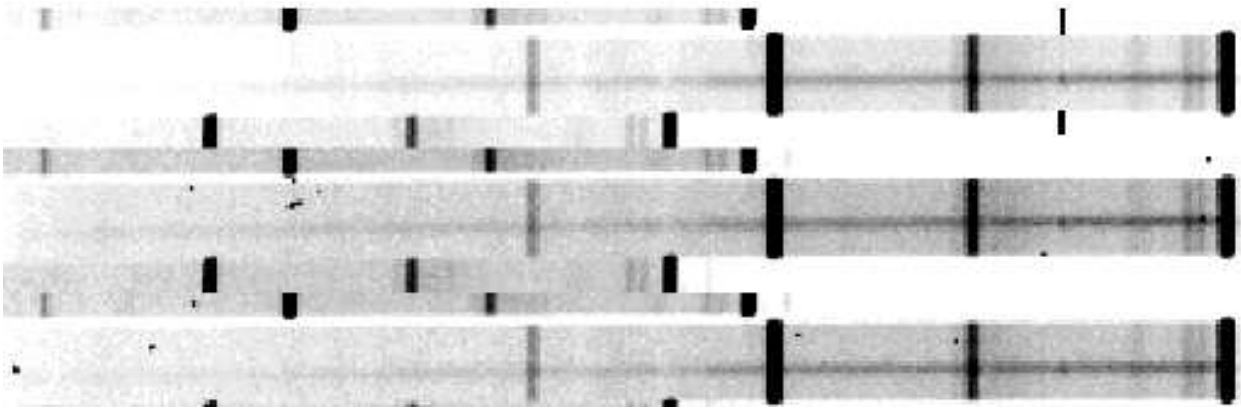}
\caption[]{
\singlespace
\small
Two-dimensional spectra for 2148.130358 (CNOC2~J215031.8$-$053504).
At top is the coadded cosmic-ray-cleaned 2-D spectrum of the object from one
mask, and below that are the two raw 2-D spectra.
Blue wavelengths are at left and red at right,
and strong night sky lines at 5577~\AA\ and 5892~\AA\ are visible.
Also visible are parts of the 2-D spectra of neighboring slits above and below
the object's 2-D spectra.
The discrepant line is between the 5892~\AA\ and 6300~\AA\ night sky lines,
marked by two line segments bracketing the coadded 2-D spectrum.
It is the only emission line in this spectrum.
}\label{2148c1A072}
\end{figure}

We also discuss the spectra and morphology of each candidate lens.
The seeing in the relevant direct and spectroscopic images 
is between 0\farcs8 and 1\farcs1.
However, the MOS imaging has a coarse pixel scale 
and significant defocusing across the chip, resulting in image
quality and resolution that is not particularly good in general.
Where possible, we instead discuss the morphology of our lens candidates on
higher quality {\em V}, {\em R}, and {\em z\arcmin} images obtained using
the CFH12k mosaic CCD camera on CFHT
as part of the Red-Sequence Cluster Survey \citep{gy00a,gea00}.
These images are at least a magnitude deeper than the CNOC2 imaging,
have better seeing (0\farcs6$-$0\farcs8 FWHM),
and are better sampled (0\farcs2/pixel).
We also make use of $R$ data obtained at the WHT \citep{hea00}
which is deeper than the MOS data but of comparable seeing and sampling.
Figure~\ref{figure} shows $R$ images of the the six lensing galaxies
(all from CFH12k except for the image of 0920.180194).
Figure~\ref{2148c1A072} shows an example of our two-dimensional spectra.

\subsubsection{0920.180194 (CNOC2 J092123.3+363613)}	\label{0920.180194}
\paragraph{Morphology and Spectrum:}
This candidate lens appears slightly asymmetric in the direct MOS images.
The deflector galaxy redshift is secure from \oii\ emission plus CN and Ca~H+K
absorption.
In the spectral image, the discrepant emission line is less spatially extended
than the \oii\ line, but about as extended as the continuum emission.
Spectra for this object were taken through only one mask.
The apparent broad line in the blue is a flatfield artifact.

\paragraph{Discrepant Line Identification:}
The discrepant line is {\em either lensed \lya\ or projected} \oii,
but probably the former (see the first paragraph of section \S\ref{Notes}).
Lensed \oii\ can be ruled out because if the discrepant line is in fact \OII, 
its redshift is lower than that of the CNOC2 target galaxy 
whose spectral features dominate the observed spectrum.
{\em Projected \lya\ can be ruled out in this and all other cases of potential
Ly$\alpha$ emission except 2148.150598,}
since discrepant line emission is seen at a projected spatial separation of
less than $\theta_E$ from the center of the putative deflector galaxy continuum.
For \lya\ emission to be seen at this observed position, it must be lensed 
\citep{bn92}.

\subsubsection{0223.191225 (CNOC2 J022336.4$-$000602)}	\label{0223.191225}
\paragraph{Morphology and Spectrum:}
\oii\ emission plus Ca~H and H$\delta$ absorption provide a secure deflector
galaxy redshift for this target.  The discrepant emission line
is present in spectra obtained with two different masks on two different nights.
Note that the wavelengths of the two emission lines visible in the spectrum
(Figure~\ref{stack}) match the expected ratio for
[N\,{\sc i}]\,$\lambda$5199 and [He\,{\sc ii}]\,$\lambda$4686 at $z=0.08931$,
but this AGN identification is extremely unlikely since no \OIII\ or H$\beta$
is seen when they should be at least 10 times stronger \citep{ost89},
and since it does not explain the observed Ca~H and H$\delta$ absorption.
As with 0920.180194, if the discrepant emission line is in fact \OII\ its
redshift is lower than that of the CNOC2 target galaxy whose spectral features
dominate the observed spectrum, and thus this system would have
to be a superposition rather than a lens.  
The CFH12k {\em Rz\arcmin} images of this field (\S\ref{Notes})
show that this is a
disk galaxy with a prominent bulge.  The galaxy is asymmetric in the $R$ image:
the disk emission is stronger to the W of the bulge than the E and the bulge
appears offset to the N of the disk, probably due to an inclination effect.
Since the discrepant line lies blueward of the $R$ band,
this asymmetry cannot be identified as arising from it alone (and vice versa).

\paragraph{Discrepant Line Identification:}
The discrepant line is {\em either lensed \lya\ or projected} \oii\ 
but probably the former (see the first paragraph of this section).
As with 0920.180194, lensed \oii\ can be ruled out because if the discrepant 
line is \oii\ it is from a galaxy with a lower redshift than the CNOC2 target 
galaxy.

\subsubsection{1447.111371 (CNOC2 J144958.6+085447)}	\label{1447.111371}
\paragraph{Morphology and Spectrum:}
This candidate lens has a close neighbor galaxy of unknown redshift 
located 2\farcs2 to the W and 0\farcs9 to the N.
The deflector galaxy redshift is secure from
\oii\ emission plus CN, H8, and Ca~H+K absorption.
The discrepant emission line is present in spectra obtained with four different
masks on two different nights.
There appears to be a variation in the strength of the line relative to the
\oii\ emission line and the continuum in one of the observations, but
	a ratio of that spectrum to the others shows that 
the variation is within the expected noise level.
If the discrepant line is \oii,
the velocity separation from the CNOC2 target galaxy would be $\sim$4000\,\kms.
On the CFH12k {\em VRz\arcmin} images of this field (\S\ref{Notes}),
the candidate appears elongated E-W in the nuclear regions,
somewhat different from the NE-SW major axis of the outer regions.  
This is confirmed by the WHT $R$ images (\S\ref{Notes}).
The elongation may be intrinsic (e.g. a bar)
or from a very close ($<$1\arcsec) projected neighbor.

\paragraph{Discrepant Line Identification:}
If the discrepant line is \oii, it is almost certainly
not from a lensed galaxy since $\theta_E=0\farcs06$ in that case.
Thus the line is {\em either lensed \lya\ or projected} \oii.

\subsubsection{2148.130358 (CNOC2 J215031.8$-$053504)}	\label{2148.130358}
\paragraph{Morphology and Spectrum:}
On the CFH12k {\em VRz\arcmin} images of this field (\S\ref{Notes}),
this candidate lens is circularly symmetric with
no signs of any morphological peculiarity.
The deflector galaxy redshift is secure from Ca~H+K and G-band absorption.
The discrepant emission line is present in spectra obtained with two different
masks on two different nights.
Its strength relative to the continuum is the same in both.  

\paragraph{Discrepant Line Identification:}
The estimated $\theta_E$ for this system is large enough that
{\em the discrepant line must be from a lensed galaxy}
regardless of whether it is \lya\ or \oii.

\subsubsection{0223.110191 (CNOC2 J022552.1$-$000018)}	\label{0223.110191}
\paragraph{Morphology and Spectrum:}
This candidate lens was identified in the CNOC2 red spectroscopy followup.
The discrepant line has the highest flux and equivalent width in our sample.
The deflector galaxy redshift was originally obtained only from \oii\ but is
confirmed by H$\beta$, \OIII, and H$\alpha$ in the followup spectroscopy.
The discrepant line is confirmed to be \oii\ at $z=0.806$ by the detection of
H$\beta$ at the same redshift (visible in Figure~\ref{stack2},
redward of H$\alpha$ from the CNOC2 target galaxy).
On the CFH12k {\em Rz\arcmin} images of this field (\S\ref{Notes}), the object
is compact but nonetheless more extended in the NNE-SSW direction than other
nearby compact objects and probable point sources.  The object is too compact
to draw any further conclusions.

\paragraph{Discrepant Line Identification:}
Since lensing probability goes as $\theta_E^2$, 
the small value of $\theta_E$ 
means that the discrepant line is
{\em probably from a projected galaxy} rather than a lensed galaxy.
It is also worth noting that the high REW of the discrepant \oii\ line suggests
that the $z=0.806$ system contributes half of the observed luminosity even if
its \oii\ REW is 100\,\AA, the maximum typically seen (\S\ref{Projections}).

\subsubsection{2148.150598 (CNOC2 J215057.9$-$055139)}	\label{2148.150598}
\paragraph{Morphology and Spectrum:}
On the CFH12k {\em VRz\arcmin} images of this field (\S\ref{Notes}), this faint
galaxy is slightly asymmetrically extended to the SE at essentially all
isophotes.  This is confirmed by the WHT $R$ images (\S\ref{Notes}).
From visual inspection of the available images, however, it does subjectively
appear to be a single galaxy rather than a close projection.
The deflector galaxy redshift is less secure than the others, being based on
strong \OIII\,$\lambda$5007 plus weak \OIII\,$\lambda$4959 and H$\beta$
emission, but the $R_{cor}$ value of the 
cross-correlation is above the cutoff for CNOC2 emission line galaxies.
Also, the two strong lines do not match any pair of known emission lines.
The discrepant emission line is present in spectra obtained with two different
masks (but on the same night).
The relative strengths of the two observed emission lines
agree very well between the two observations.
The apparent line at 4444~\AA\ (observed) is spurious.

\paragraph{Discrepant Line Identification:}
Given the small $\theta_E$ estimated for this object,
the discrepant line here is very probably either 
Ly$\alpha$ from a projected $z=3.97$ galaxy
or \oii\ from a projected $z=0.62$ galaxy.
The first possibility is unlikely:
the projected galaxy is estimated to have $I\sim23.2$,
at which the surface density of $z\simeq4$ Lyman-break galaxies is only
$\sim$14~deg$^{-2}$,
of which only $\sim$50\% have Ly$\alpha$ in emission \citep{ste99}.
The line flux of 5.4 10$^{-17}$ ergs cm$^{-2}$ s$^{-1}$ would also be
rather high for an unlensed $z=3.97$ galaxy.  
\citet{sea99} do find 3 galaxies of comparable Ly$\alpha$ flux in a 78~arcmin$^2$
narrowband survey, but the redshift targeted was one at which the observed field
was known to be overdense in Lyman-break galaxies by a factor $\sim$6.
Thus we conclude that the discrepant emission line in 2148.150598
is {\em most likely projected} \oii\ {\em at $z=0.62$.}
The slightly asymmetric morphology of this source
lends {\em a posteriori} support to this conclusion,
although {\em a priori} we did not believe it indicated a projected galaxy.

\subsection{Projections or Lenses?} 	\label{Possibilities}

Estimating a robust expected number of lenses in the CNOC2 survey requires
a complicated calculation that would produce only an uncertain estimate.
This is because the magnification of a lens 
(and thus the surface density of the faint galaxy population which could be
  responsible for our candidate lenses) depends very sensitively on the
impact parameter between the source and deflector galaxies.
Our observed lensing frequency from the original CNOC2 survey
is $\sim$1 good lens candidate per 1500$-$2000 galaxies to $R\lesssim21.5$.  
This is a lower limit due to limited line detectability and redshift coverage.
The correction for limited line detectability will be at least a factor of two,
since only $\sim$50\% of $z\simeq3$ galaxies show \lya\ in emission
\citep{sea99}.
We ignore lensed $z\lesssim1$ galaxies since higher-$z$ lenses dominate the
  lensing cross-section (cf. the values of $\theta_E$(\oii) and
  $\theta_E$(Ly$\alpha$) in Table~\ref{tab_phot}).
The correction for limited redshift coverage is complicated since the source
magnitude and the lensing cross-section are correlated with the source redshift,
but it is also likely to be at least a factor of two for the following reason.  
CNOC2 is sensitive to source galaxies only over a redshift range of 
$\Delta z = 1.55$ ($2.62 < z < 4.17$), 
again ignoring the less common $z\lesssim1$ lenses.
There is an adjacent redshift range of similar $\Delta z$ at lower redshifts
where CNOC2 should have detected a similar number of lenses if it had had the
appropriate spectral coverage, since our observed candidate deflector galaxies
have roughly the same Einstein radii for source galaxies at such redshifts.
In other words, the lensing probabilities for source galaxies at those lower
redshifts --- whose line emission would be detectable in data of CNOC2 quality,
unlike emission from objects at redshifts higher than those to which CNOC2
is sensitive, which is likely to be too faint to be detected --- 
are roughly the same as for the redshifts observable in CNOC2.
Our resulting estimated lens frequency of 1 in $\lesssim$~400$-$500
is consistent with the approximately 1 in 600 frequency of radio rings
in the MGV survey \citep{mel92}.

\section{Conclusion}	\label{Conclusion}

In the course of the CNOC2 field galaxy redshift survey and subsequent followup,
we have identified six spectroscopically-selected gravitational lens candidates.
Each 
exhibits a firm redshift from multiple
spectral features, plus a discrepant emission line.  Ly$\alpha$ or \OII\ are
the only reasonable identifications for the discrepant lines.
Based on the estimated Einstein radii of the putative deflector galaxies,
four of the discrepant lines are more likely to be lensed Ly$\alpha$
than lensed \oii,
and the remaining two are unlikely to be lensed Ly$\alpha$ or lensed \oii\ and
are therefore probably \oii\ from projected $z<1$ galaxies.  In one case the
\oii\ identification is confirmed by the detection of H$\beta$.
We estimate that there should be only $1.9\pm0.7$ chance projections
of objects which could explain the observed discrepant emission lines in the
original CNOC2 sample.  
Thus there is only a 3\% chance that all five of the original CNOC2 survey
candidates are projections, and it is likely that
three or four of our six candidates are real gravitational lenses.

Galaxy-galaxy lenses can be used to
constrain the mass distributions of the deflector galaxies.  
While our sample is small, only two such systems have previously been reported
\citep{war96,hew99}.  Unlike \citet{war96}, we do not preselect for red,
early-type deflector galaxies and thus the two samples may probe the masses of
different galaxy types.  In addition, more galaxy-galaxy lenses are needed to
search for rare but valuable cases where two or more galaxies at different
redshifts are lensed by the same object.  Such systems may be able to 
simultaneously constrain $\Omega$ and $\Lambda$ since all lensed sources must
share the same relation between angular size distance and redshift \citep{lp98}.

Further observations of these systems are therefore warranted.
High spatial resolution narrow-band imaging can provide the surface brightness
distributions of the lensed emission, needed to constrain the deflector galaxy
mass distributions.
Forthcoming integral field spectroscopy should 
confirm the discrepant 
lines, determine the emission line region morphologies,
estimate the velocity dispersion of the deflector galaxies, 
reveal if the discrepant emission lines have the redward-asymmetric profile
common to Ly$\alpha$ at high redshift,
and possibly detect other emission or absorption lines from the source galaxies
such as H$\beta$/\OIII\ at low redshift or \civ\ at high redshift.
For high-$z$ source galaxies, near-IR spectroscopy can study the rest-frame
optical emission lines to constrain masses from the observed linewidths,
to compare star formation rate estimates from Ly$\alpha$, \oii\, and H$\beta$,
and to compare line ratios with star-forming galaxies at lower redshifts.

\acknowledgements
We thank CTAC and the CFHT for generous allocations of telescope time, 
the CFHT operators for their dedicated assistance during observing,
D. Balam for assistance with the astrometry, C. Burns for helpful discussions,
H. Hoekstra for the use of WHT images of some CNOC2 patches,
and the referee for helpful comments.
CNOC was supported by a Collaborative Program grant from NSERC,
as well as by individual NSERC operating grants to HY and RC.
HL acknowledges support provided by NASA through Hubble Fellowship grant
\#HF-01110.01-98A awarded by the Space Telescope Science Institute, which
is operated by the Association of Universities for Research in Astronomy,
Inc., for NASA under contract NAS 5-26555.

\footnotesize

\begin{thebibliography}{30}
\expandafter\ifx\csname natexlab\endcsname\relax\def\natexlab#1{#1}\fi

\bibitem[{{Blandford} \& {Narayan}(1992)}]{bn92}
{Blandford}, R.~D. \& {Narayan}, R. 1992, \araa, 30, 311

\bibitem[{{Bruzual A.} \& {Charlot}(1996)}]{bc96}
{Bruzual A.}, G. \& {Charlot}, S. 1996, preprint

\bibitem[{{Coleman} {et~al.}(1980){Coleman}, {Wu}, \& {Weedman}}]{cww80}
{Coleman}, G.~D., {Wu}, C.-C., \& {Weedman}, D.~W. 1980, \apjs, 43, 393

\bibitem[{{Gladders} \& {Yee}(2000{\natexlab{a}})}]{gy00c}
{Gladders}, M.~D. \& {Yee}, H. K.~C. 2000{\natexlab{a}}, in preparation

\bibitem[{{Gladders} \& {Yee}(2000{\natexlab{b}})}]{gy00a}
---. 2000{\natexlab{b}}, to appear in ``Cosmic Evolution and Galaxy Formation:
  Structure, Interactions and Feedback" (astro-ph/0002340)

\bibitem[{{Gladders} {et~al.}(2000){Gladders}, {Yee}, \& {Hall}}]{gea00}
{Gladders}, M.~D., {Yee}, H. K.~C., \& {Hall}, P.~B. 2000, in preparation

\bibitem[{{Hewett} {et~al.}(1999){Hewett}, {Warren}, {Willis},
  {Bland-Hawthorn}, \& {Lewis}}]{hew99}
{Hewett}, P.~C., {Warren}, S.~J., {Willis}, J.~P., {Bland-Hawthorn}, J., \&
  {Lewis}, G.~F. 1999, in to appear in ``Imaging the Universe in Three
  Dimensions'', ed. W. van Breugel and J. Bland-Hawthorn (astro-ph/9905316)

\bibitem[{{Hoekstra et al.}(2000)}]{hea00}
{Hoekstra et al.}, H. 2000, in preparation

\bibitem[{{Hogg} {et~al.}(1998){Hogg}, {Cohen}, {Blandford}, \&
  {Pahre}}]{hog98}
{Hogg}, D.~W., {Cohen}, J.~G., {Blandford}, R., \& {Pahre}, M.~A. 1998, \apj,
  504, 622

\bibitem[{{Kochanek}(1994)}]{koc94}
{Kochanek}, C.~S. 1994, \apj, 436, 56

\bibitem[{{Kochanek}(1995)}]{koc95}
---. 1995, \apj, 445, 559

\bibitem[{{Kochanek} {et~al.}(2000{\natexlab{a}}){Kochanek}, {Falco}, {Impey},
  {Lehar}, {McLeod}, {Rix}, {Keeton}, {Munoz}, \& {Peng}}]{koc98}
{Kochanek}, C.~S., {Falco}, E.~E., {Impey}, C.~D., {Lehar}, J., {McLeod},
  B.~A., {Rix}, H.~W., {Keeton}, C.~R., {Munoz}, J.~A., \& {Peng}, C.~Y.
  2000{\natexlab{a}}, \apj, 535, 692

\bibitem[{{Kochanek} {et~al.}(2000{\natexlab{b}}){Kochanek}, {Falco}, {Impey},
  {Lehar}, {McLeod}, {Rix}, {Keeton}, {Munoz}, \& {Peng}}]{koc99}
---. 2000{\natexlab{b}}, \apj, in press (astro-ph/9909018)

\bibitem[{{Lin} {et~al.}(1999){Lin}, {Yee}, {Carlberg}, {Morris}, {Sawicki},
  {Patton}, {Wirth}, \& {Shepherd}}]{cnoc2.1}
{Lin}, H., {Yee}, H. K.~C., {Carlberg}, R.~G., {Morris}, S.~L., {Sawicki}, M.,
  {Patton}, D.~R., {Wirth}, G.~D., \& {Shepherd}, C.~W. 1999, \apj, 518, 533

\bibitem[{{Lin} {et~al.}(2000){Lin}, {Yee}, {Carlberg}, {Morris}, {Sawicki},
  {Patton}, {Wirth}, \& {Shepherd}}]{lea00}
---. 2000, in preparation

\bibitem[{{Link} \& {Pierce}(1998)}]{lp98}
{Link}, R. \& {Pierce}, M.~J. 1998, \apj, 502, 63

\bibitem[{{Miralda-Escude} \& {Lehar}(1992)}]{mel92}
{Miralda-Escude}, J. \& {Lehar}, J. 1992, \mnras, 259, 31P

\bibitem[{{Morris} {et~al.}(2000){Morris}, {Lin}, {Carlberg}, {Sawicki}, \&
  {Hall}}]{mor00}
{Morris}, S.~L., {Lin}, H., {Carlberg}, R.~G., {Sawicki}, M., \& {Hall}, P.
  2000, in preparation

\bibitem[{{Osterbrock}(1989)}]{ost89}
{Osterbrock}, D.~E. 1989, Astrophysics of Gaseous Nebulae and Active Galactic
  Nuclei (Mill Valley: University Science Books), 357

\bibitem[{{Ratnatunga} {et~al.}(1999){Ratnatunga}, {Griffiths}, \&
  {Ostrander}}]{rat99}
{Ratnatunga}, K.~U., {Griffiths}, R.~E., \& {Ostrander}, E.~J. 1999, \aj, 117,
  2010

\bibitem[{{Sandage}(1978)}]{san78}
{Sandage}, A. 1978, \aj, 83, 904

\bibitem[{{Schade} {et~al.}(1999){Schade}, {Lilly}, {Crampton}, {Ellis}, {Le
  Fevre}, {Hammer}, {Brinchmann}, {Abraham}, {Colless}, {Glazebrook}, {Tresse},
  \& {Broadhurst}}]{sch99}
{Schade}, D., {Lilly}, S., {Crampton}, D., {Ellis}, R., {Le Fevre}, O.,
  {Hammer}, F., {Brinchmann}, J., {Abraham}, R., {Colless}, M., {Glazebrook},
  K., {Tresse}, L., \& {Broadhurst}, T. 1999, \apj, 525, 31

\bibitem[{{Steidel} {et~al.}(1999{\natexlab{a}}){Steidel}, {Adelberger},
  {Giavalisco}, {Dickinson}, \& {Pettini}}]{ste99}
{Steidel}, C.~C., {Adelberger}, K.~L., {Giavalisco}, M., {Dickinson}, M., \&
  {Pettini}, M. 1999{\natexlab{a}}, \apj, 519, 1

\bibitem[{{Steidel} {et~al.}(1999{\natexlab{b}}){Steidel}, {Adelberger},
  {Shapley}, {Pettini}, {Dickinson}, \& {Giavalisco}}]{sea99}
{Steidel}, C.~C., {Adelberger}, K.~L., {Shapley}, A.~E., {Pettini}, M.,
  {Dickinson}, M., \& {Giavalisco}, M. 1999{\natexlab{b}}, \apj, 532, 170

\bibitem[{{Stern} {et~al.}(2000){Stern}, {Bunker}, {Spinrad}, \& {Dey}}]{ste00}
{Stern}, D., {Bunker}, A.~J., {Spinrad}, H., \& {Dey}, A. 2000, \apj, in press
  (astro-ph/0002239)

\bibitem[{{Tonry} \& {Kochanek}(2000)}]{tk00}
{Tonry}, J.~L. \& {Kochanek}, C.~S. 2000, \aj, 119, 1078

\bibitem[{{Warren} {et~al.}(1996){Warren}, {Hewett}, {Lewis}, {M\o ller},
  {Iovino}, \& {Shaver}}]{war96}
{Warren}, S.~J., {Hewett}, P.~C., {Lewis}, G.~F., {M\o ller}, P., {Iovino}, A.,
  \& {Shaver}, P.~A. 1996, \mnras, 278, 139

\bibitem[{{Warren} {et~al.}(1999){Warren}, {Lewis}, {Hewett}, {M\o ller},
  {Shaver}, \& {Iovino}}]{war99}
{Warren}, S.~J., {Lewis}, G.~F., {Hewett}, P.~C., {M\o ller}, P., {Shaver}, P.,
  \& {Iovino}, A. 1999, \aap, 343, L35

\bibitem[{{Yee} {et~al.}(1996){Yee}, {Ellingson}, \& {Carlberg}}]{yec96}
{Yee}, H. K.~C., {Ellingson}, E., \& {Carlberg}, R.~G. 1996, \apjs, 102, 269

\bibitem[{{Yee} {et~al.}(2000){Yee}, {Morris}, {Lin}, {Carlberg}, {Hall},
  {Sawicki}, {Patton}, {Wirth}, {Ellingson}, \& {Shepherd}}]{yea00}
{Yee}, H. K.~C., {Morris}, S.~L., {Lin}, H., {Carlberg}, R.~G., {Hall}, P.~B.,
  {Sawicki}, M., {Patton}, D.~R., {Wirth}, G.~D., {Ellingson}, E., \&
  {Shepherd}, C.~W. 2000, \apjs, 131, in press (astro-ph/0004026)

\end{thebibliography}

\clearpage
\begin{deluxetable}{cccccccccclrc} 
\tablecaption{CNOC2 Candidate Spectroscopic Gravitational Lenses\label{tab_lenses}}
\rotate
\tiny
\tablewidth{692.93245pt}
\tablehead{
\colhead{}   & \colhead{}                    & \colhead{}           & \colhead{}          & \colhead{}    & \colhead{}                 & \colhead{}                 & \colhead{}           & \colhead{Discrepant}                 & \colhead{Discrepant}                   & \colhead{}     & \colhead{}      & \colhead{IAU}          \\[.2ex]
\colhead{}   & \colhead{}                    & \colhead{}           & \colhead{Peak}      & \colhead{}    & \colhead{$z_{\rm source}$} & \colhead{$z_{\rm source}$} & \colhead{}           & \colhead{Line}                       & \colhead{Line Rest $\lambda$}          & \colhead{Line} & \colhead{EW}    & \colhead{Nomenclature} \\[.2ex]
\colhead{ID} & \colhead{$z_{\rm deflector}$} & \colhead{$\sigma_z$} & \colhead{$R_{cor}$} & \colhead{Scl} & \colhead{(if Ly$\alpha$)}  & \colhead{(if O\,{\sc ii})} & \colhead{$\sigma_z$} & \colhead{Observed $\lambda$}         & \colhead{at $z_{\rm deflector}$}       & \colhead{Flux} & \colhead{(\AA)} & \colhead{Object Name}
}
\startdata
0920.180194 & 0.56502 & 25 & 4.67 & 4                  & 3.26302 & 0.39051 & 25 & 5183 & 3313 & 0.60$\pm$0.05 & 18.5 & J092123.3$+$363613 \\ 
0223.191225 & 0.51142 & 18 & 5.06 & 5\tablenotemark{a} & 3.17764 & 0.36266 & 30 & 5077 & 3558 & 1.20$\pm$0.12 & 20.1 & J022336.4$-$000602 \\ 
1447.111371 & 0.39545 & 18 & 5.21 & 5\tablenotemark{b} & 3.33820 & 0.41503 & 42 & 5274 & 3778 & 1.30$\pm$0.08 & 14.5 & J144958.6$+$085447 \\ 
2148.130358 & 0.37396 & 19 & 6.08 & 2                  & 3.96494 & 0.61946 & 12 & 6037 & 4395 & 2.47$\pm$0.05 & 10.5 & J215031.8$-$053504 \\ 
0223.110191 & 0.30311 & 20 & 11.26 & 5                 & \nodata & 0.80581 & 12 & 6731 & 5163 & 4.49$\pm$0.15 & 55.8 & J022552.1$-$000018 \\ 
2148.150598 & 0.10589 & 17 & 3.26 & 5                  & 3.96509 & 0.61951 & 12 & 6038 & 5457 & 0.54$\pm$0.06 &  9.0 & J215057.9$-$055139 \\ 
\enddata
\tablenotetext{a}{Classified as Scl=4 in one of two independent spectra.}
\tablenotetext{b}{Classified as Scl=4 in one of four independent spectra.}
\tablecomments{
Each candidate's ID number is its CNOC2 catalog number, consisting of a
four-digit patch code plus a six-digit field+object code after a decimal point.
All redshifts are measured from cross-correlation on the 
average of all available individual spectra.  
Redshift uncertainties for $z_{\rm deflector}$ and $z_{\rm source}$ 
(whether Ly$\alpha$ or [O\,{\sc ii}]) are given in units of 0.00001 in $z$.
$R_{cor}$ values are explained in \S\ref{Selection}.
Spectral classes (Scl) are 2 for absorption lines, 4 for emission+absorption,
and 5 for emission-dominated.  In cases of multiple spectra, spectral classes
are taken from the spectrum with the highest $R_{cor}$.
Wavelengths of the discrepant lines are given in \AA.
Line fluxes are in cgs units of 10$^{-16}$ ergs cm$^{-2}$ s$^{-1}$.
Equivalent widths (EW) are measured in the observed frame
relative to the putative deflector galaxy continuum emission.
The IAU nomenclature object names provide RA and DEC in the J2000 system,
and should be preceded by the acronym CNOC2, e.g. CNOC2~J092123.3$+$363613.
}
\end{deluxetable}

\begin{deluxetable}{ccccccccccccccc} 
\tablecaption{CNOC2 Candidate Lenses: Photometry and Derived Parameters\label{tab_phot}}
\rotate
\tiny
\tablewidth{599.81715pt}
\tablehead{
\colhead{ID} & \colhead{U} & \colhead{err} & \colhead{B} & \colhead{err} & \colhead{V} & \colhead{err} & \colhead{R} & \colhead{err} & \colhead{I} & \colhead{err} & \colhead{SED} & \colhead{$M_B$} & \colhead{$\theta_E$(Ly$\alpha$)} & \colhead{$\theta_E$(\oii)} }
\startdata
0920.180194 & 21.46 & 0.09 & 21.64 & 0.05 & 20.82 & 0.04 & 19.89 & 0.04 & 19.16 & 0.04 & 1.71 & $-$21.11 & 1\farcs02 & \nodata   \\ 
0223.191225 & 22.61 & 0.17 & 22.46 & 0.09 & 21.30 & 0.04 & 20.22 & 0.04 & 19.44 & 0.04 & 0.91 & $-$20.51 & 0\farcs87 & \nodata   \\ 
1447.111371 & 21.48 & 0.11 & 21.52 & 0.05 & 20.66 & 0.04 & 19.94 & 0.04 & 19.33 & 0.04 & 2.01 & $-$20.04 & 0\farcs76 & 0\farcs04 \\ 
2148.130358 & 21.74 & 0.11 & 21.69 & 0.04 & 20.25 & 0.04 & 19.15 & 0.03 & 18.40 & 0.04 & 0.38 & $-$20.51 & 1\farcs13 & 0\farcs52 \\ 
0223.110191 & 22.09 & 0.14 & 22.42 & 0.07 & 21.70 & 0.05 & 21.02 & 0.07 & 20.32 & 0.07 & 2.08 & $-$18.30 & \nodata   & 0\farcs29 \\ 
2148.150598 & 21.69 & 0.12 & 22.10 & 0.05 & 21.52 & 0.07 & 21.36 & 0.05 & 20.72 & 0.10 & 3.31 & $-$15.77 & 0\farcs15 & 0\farcs13 \\ 
\enddata
\tablecomments{
Each candidate's ID number is its CNOC2 catalog number, consisting of a
four-digit patch code plus a six-digit field+object code after a decimal point.
SED classes are derived by fitting spectral templates to the {\em UBVRI}
photometry \citep{yea00}.  All values between $-$0.5 and 4.5 are allowed, with 
0, 1, 2, 3 being \cite{cww80} E/S0, Sbc, Scd, and Im templates respectively,
and 4 being a very blue vigorously star-forming GISSEL template \citep{bc96}.
$M_B$ values are calculated using these SED classes, normalizing to all
the available photometry instead of only one band \citep{cnoc2.1}.
$\theta_E$(Ly$\alpha$) is the estimated Einstein radius if the putative lensed
emission line is Ly$\alpha$, and $\theta_E$(\oii) is the estimated Einstein
radius if the putative lensed emission line is \OII\ (see \S\ref{Einstein}).
No $\theta_E$(\oii) is calculated for 0920.180194 or 0223.191225 because
the discrepant line lies blueward of the deflector galaxy \oii\ line, so if the
discrepant line is \oii\ in those cases it must arise from a foreground galaxy.
No $\theta_E$(Ly$\alpha$) is calculated for 0223.110191 because its discrepant
line is confirmed to be \oii\ by the detection of H$\beta$ at the same redshift.
}
\end{deluxetable}

%
%
%

\end{document}